\documentclass[amsmath,11pt]{article}

\usepackage{amssymb,amssymb,epsfig,bm}

\date{\empty}

\begin{document}

\title{\bf The peculiar Jeans length}
\author{Christos G. Tsagas\\ {\small Section of Astrophysics,
Astronomy and Mechanics, Department of Physics}\\ {\small Aristotle
University of Thessaloniki, Thessaloniki 54124, Greece}\\ \textit{\small and}\\ {\small Clare Hall, University of Cambridge, Herschel Road, Cambridge CB3 9AL, UK.}}

\maketitle

\begin{abstract}
Typical observers in the universe do not follow the smooth Hubble expansion, but move relative to it. Such bulk peculiar motions introduce a characteristic scale that is closely analogous to the familiar Jeans length. This ``peculiar Jeans length'' marks the threshold below which relative-motion effects dominate the linear kinematics. There, cosmological measurements can vary considerably between the bulk-flow frame and that of the Hubble expansion, entirely due to the observers' relative motion. When dealing with the deceleration parameter, we find that the peculiar Jeans length varies between few and several hundred Mpc. On these scales, the deceleration parameter measured by the bulk-flow observers can be considerably larger (or smaller) than its Hubble-frame counterpart. This depends on whether the peculiar motion is locally expanding (or contracting), relative to the background expansion. Then, provided expanding and contracting bulk flows are randomly distributed, nearly half of the observers in the universe could be misled to think that their cosmos is over-decelerated. The rest of them, on the other hand, may come to believe that their universe is under-decelerated, or even accelerated in some cases. We make two phenomenological predictions that could in principle support this scenario.
\end{abstract}

\section{Introduction}\label{sI}
Surveys of peculiar-velocity fields have repeatedly reported the existence of large-scale bulk flows, moving coherently with respect to the mean universal expansion.\footnote{We assume that the Hubble flow selects the frame, relative to which peculiar velocities can be defined and measured. In what follows, we will use the terms ``reference frame'', or ``Hubble frame'', when referring to this idealised coordinate system.} The size of these motions is of the order of few to several hundred Mpc and their velocities vary between few and several hundred km/sec (e.g.~see~\cite{Aetal} and references therein). Our galaxy and the nearby Local Group, for example are ``drifting'' at approximately 600~km/sec with respect to the smooth Hubble flow. Such large-scale peculiar motions are believed to be a relatively recent addition to the phenomenology of our universe and the inevitable result of the ongoing structure-formation process.

In relativity, moving observers generally experience different versions of what one might call ``reality'', entirely due to their relative motion. Then, in principle at least, observers living in a typical galaxy (like our Milky Way) may ``see'' a different universe than those following the smooth Hubble expansion. Such an apparent effect should be local of course, since peculiar velocities are expected to fade away as we move on to progressively larger lengths. Nevertheless, to an unsuspecting observer, local events may appear as recent global ones, provided the scales involved are large enough (of the order of few hundred Mpc). With these thoughts in mind, we apply linear relativistic cosmological perturbation theory to investigate the implications of large-scale peculiar velocity fields for the interpretation of key cosmological parameters. Assuming a Friedmann-Robertson-Walker (FRW) background filled with pressure-free dust, we find that the associated relative-motion effects have a particular scale-dependence. More specifically, we show that bulk peculiar motions introduce a characteristic length scale, which depends on the velocity of the moving domain and is closely analogous to the familiar (from standard linear perturbation theory) Jeans length. The latter has been known to mark the threshold below which pressure gradients dominate over the background gravity and thus dictate the evolution of linear density perturbations. In analogy, the \textit{peculiar Jeans length} reported here sets the scale below which the linear kinematics of the bulk-flow observers are dominated by relative-motion effects, rather than by the background Hubble expansion. Such kinematic ``contamination'' can seriously interfere with the way these observers interpret their cosmological data, given that the typical sizes of the affected regions are between few to several hundred Mpc.

One of the affected quantities is the deceleration parameter of the universe. We find that on scales smaller than the aforementioned peculiar Jeans length, the measurements of the deceleration parameter can be significantly contaminated by relative-motion effects. In particular, the deceleration parameter measured in the rest-frame of (slightly) expanding bulk flows can be considerably larger than its Hubble-frame counterpart. Inside (slightly) contracting peculiar motions, on the other hand, the effect is reversed. There, even the sign of the deceleration parameter can change from positive to negative. Although these are purely local relative-motion effects, the affected scales are typically large enough to make them appear as global. Assuming that there is no generic bias for expanding (or contracting) bulk flows on cosmological scales (i.e.~on lengths larger than 100~Mpc), the chances of residing in one of them should be approximately 50\%. Then, roughly half of the observers living in a nearly-flat, almost-FRW universe (filled with ordinary dust) may think that their cosmos is over-decelerated, while the rest could be misled to believe that their universe is under-decelerated, or even accelerated in some cases.

One might wonder whether there is a way for these unsuspecting observers to find out that they have been merely experiencing an illusion. The answer should be in the data, where one should look for the characteristic features/signatures of the bulk-flow scenario. Such a feature is the scale-distribution of the deceleration parameter. The latter should have a nonlinear scale-evolution, approaching its background value on large scales (away from the observer) and becoming increasingly more negative on small wavelengths (i.e.~near the observer), with a profile resembling the one depicted in Fig.~\ref{fig:qplot}. A second feature is an apparent (Doppler-like) dipole in the sky-distribution of the deceleration parameter. Put another way, to the bulk-flow observers, the universe should seem to accelerate faster in one direction in the sky and equally slower in the opposite. The existence of such an apparent dipole, which is the typical (trademark) signature of relative motion, has been proposed on theoretical grounds and its presence appears to have some observational backing as well (see \S~\ref{ssSB-FS}).

\section{The peculiar kinematics}\label{sPKs}
Our starting point is a perturbed FRW cosmology filled with pressureless dust and equipped with two families of observers. These are the (fictitious) idealised observers, following the smooth Hubble flow, and their real counterparts moving relative to the mean universal expansion.

\subsection{Relatively moving observers}\label{ssRMOs}
\begin{figure}[tbp]
\centering \vspace{5cm} \includegraphics{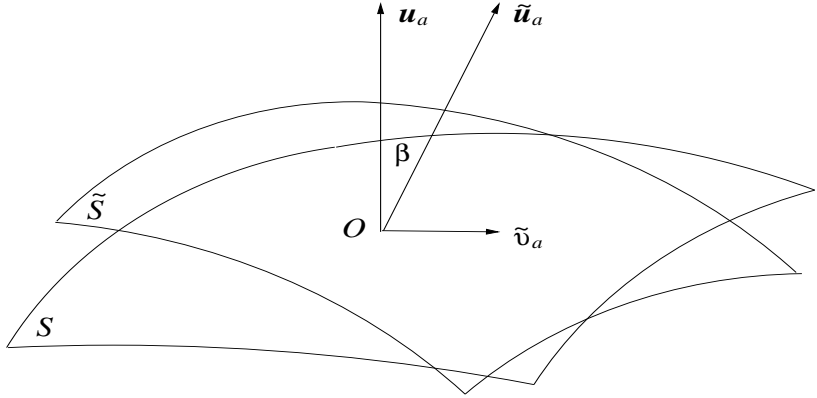} \caption{Observer ($O$) moving with peculiar velocity $\tilde{v}_a$ (where $\tilde{v}^2\ll1$ in our case), relative to the Hubble flow. The 4-velocities $u_a$ and $\tilde{u}_a$, with a hyperbolic (tilt) angle $\beta$ between them, respectively define the reference frame of the smooth universal expansion and that of the peculiar motion (see Eq.~(\ref{4vels})). The 3-D hypersurfaces $S$ and $\tilde{S}$ are normal to $u_a$ and $\tilde{u}_a$ and they respectively define the rest-spaces of the idealised observers and of their real counterparts.}  \label{fig:bflow}
\end{figure}

For non-relativistic peculiar velocities ($\tilde{v}_a$, with $\tilde{v}^2\ll1$), the aforementioned two groups of observers are related by the ``reduced'' Lorentz boost
\begin{equation}
\tilde{u}_a= u_a+ \tilde{v}_a\,,  \label{4vels}
\end{equation}
where $u_a$ and $\tilde{u}_a$ are the 4-velocities of the idealised and the real observers respectively (see Fig.~\ref{fig:bflow}).\footnote{Hereafter, tildas will denote quantities measured in the rest-frame of the bulk motion. Also Latin indices will run from 0 to 3, while Greek ones will take values between 1 and 3.} Note that $u_au^a=-1=\tilde{u}_a\tilde{u}^a$ and $u_a\tilde{v}^a=0$ by construction, while $\tilde{\gamma}= (1-\tilde{v}^2)^{-1/2}$ is the associated Lorentz boost factor. The latter also determines the (hyperbolic) ``tilt'' angle ($\beta$) between the two 4-velocity vectors, since $\cosh\beta=-u_a\tilde{u}^a=\tilde{\gamma}$ (see~\cite{KE} and also Fig.~\ref{fig:bflow}). In our case, the fact that $\tilde{v}^2\ll1$ guarantees that $\tilde{\gamma}\simeq1\simeq\cosh\beta$.

Each one of these 4-velocity fields introduces an 1+3 splitting of the spacetime into a temporal direction and a corresponding 3-dimensional space. Hereafter, we will use overdots and primes, namely $\;{}^{\cdot}=u^a\nabla_a$ and $\;{}^{\prime}=\tilde{u}^a\nabla_a$, to indicate time differentiation in the $u_a$ and the $\tilde{u}_a$ frames respectively (with $\nabla_a$ being the 4-dimensional covariant derivative operator). Also, given that $g_{ab}$ is the 4-D spacetime metric, the symmetric tensors $h_{ab}=g_{ab}+u_au_b$ and $\tilde{h}_{ab}=g_{ab}+ \tilde{u}_a\tilde{u}_b$ project onto the 3-D hypersurfaces $S$ and $\tilde{S}$, orthogonal to the $u_a$ and the $\tilde{u}_a$ fields respectively (see Fig.~\ref{fig:bflow}). Therefore, the corresponding spatial derivative operators are ${\rm D}_a= h_a{}^b\nabla_b$ and $\tilde{\rm D}_a=\tilde{h}_a{}^b\nabla_b$~\cite{TCM,EMM}.

We should point out that we will take the viewpoint of the real (the ``tilted'') observers and concentrate on the mean kinematics of their flow. Having said that, in Appendix~\ref{app2} we provide an alternative analysis that takes the perspective of an observer following the $u_a$-frame. Both approaches reach the same conclusions, in accord with the equivalence of reference frames in relativity.

\subsection{Linear relations between the two frames}\label{ssLRBTFs}
The expansion/contraction of the aforementioned two families of observers is determined by the associated volume scalars, respectively given by $\Theta={\rm D}^au_a$  and $\tilde{\Theta}= \tilde{\rm D}^a\tilde{u}_a$ (e.g.~see~\cite{TCM,EMM}). These take positive values in the case of expansion and negative when dealing with contraction. Moreover, the time-derivatives of the volume scalars decide the deceleration/accelleration of the associated expansion/contraction. Following (\ref{4vels}), the rate of the expansion/contraction, as well as the deceleration/acceleration, experienced by the aforementioned observer groups differ, even when the peculiar velocity is non-relativistic. More specifically, we have
\begin{equation}
\tilde{\Theta}= \Theta+ \tilde{\vartheta} \hspace{15mm} {\rm and} \hspace{15mm} \tilde{\Theta}^{\prime}= \dot{\Theta}+ \tilde{\vartheta}^{\prime}\,,  \label{Thetas}
\end{equation}
to linear order~\cite{M}.\footnote{Given that peculiar velocities vanish in the FRW background, the perturbed variables satisfy the criteria for linear gauge-invariance~\cite{StWa}, which frees our analysis from gauge-related ambiguities.} Note that $\tilde{\vartheta}=\tilde{\rm D}^a\tilde{v}_a$ is the volume scalar of the peculiar motion, measured in the rest frame of the bulk flow. This scalar takes positive values inside (locally) expanding bulk flows, but it turns negative inside contracting ones. Recall also that overdots and primes indicate time derivatives in the Hubble and the tilted frames respectively. According to Eqs.~(\ref{Thetas}), we have $\tilde{\Theta}\neq\Theta$ and $\tilde{\Theta}^{\prime}\neq \dot{\Theta}$, with their difference decided by the magnitude and by the sign of $\tilde{\vartheta}$ and $\tilde{\vartheta}^{\prime}$. More specifically, the relative-motion effects are determined by the dimensionless ratios $\tilde{\vartheta}/\Theta$ and $\tilde{\vartheta}^{\prime}/\dot{\Theta}$. Also note that, although the constraint $|\tilde{\vartheta}/\Theta|\ll1$ holds at all times during the linear regime, the time-derivative ratio $|\tilde{\vartheta}^{\prime}/\dot{\Theta}|$ can exceed unity even within the linear approximation.\footnote{One can easily show that $\tilde{\vartheta}^{\prime}/\dot{\Theta}= [(\tilde{\vartheta}^{\prime}/\tilde{\vartheta})/(\dot{\Theta}/\Theta))] (\tilde{\vartheta}/\Theta)$, which allows for $|\tilde{\vartheta}^{\prime}/\dot{\Theta}|>1$ while $|\tilde{\vartheta}/\Theta|\ll1$. Clearly, if $|\tilde{\vartheta}^{\prime}/\dot{\Theta}|$ remains large for long (in cosmological terms), the linear constraint will be eventually violated.}

Analogous relations also connect the rest of the kinematic and dynamic variables measured in the two frames (see~\cite{M} for details). More specifically, on our pressure-free FRW background, the linear expressions between the 4-acceleration and between the energy-flux vectors are
\begin{equation}
\tilde{A}_a= A_a+ \tilde{v}_a^{\prime}+ H\tilde{v}_a \hspace{7.5mm} {\rm and} \hspace{7.5mm} \tilde{q}_a= q_a- \rho\tilde{v}_a\,,  \label{lrs1}
\end{equation}
respectively. Note that $H$ is the Hubble parameter in the Friedmann background. There, the peculiar velocities vanish, which means that the reference and the tilted frames ($u_a$ and $\tilde{u}_a$ respectively) coincide in an exact FRW model. Assuming that the cosmic fluid is perfect in the perturbed spacetime, we may set $q_a=0$ in the Hubble frame. Also, given the absence of matter pressure, we may set $A_a=0$ at the linear level. In the tilted frame, on the other hand, both of these perturbations take nonzero values.\footnote{The reader is referred to Appendix~\ref{app2} for an alternative treatment where the frame assumptions are reversed. There, we set $\tilde{q}_a=0=\tilde{A}_a$, with $q_a,A_a\neq0$ at the same time.} More specifically, according to Eqs.~(\ref{lrs1}), observers living inside the bulk flow measure
\begin{equation}
\tilde{A}_a= \tilde{v}_a^{\prime}+ H\tilde{v}_a \hspace{7.5mm} {\rm and} \hspace{7.5mm} \tilde{q}_a= -\rho\tilde{v}_a\,,  \label{lrs2}
\end{equation}
as a result of their  peculiar motion alone. Following (\ref{lrs2}a), the worldlines of the tilted observers are no longer timelike geodesics due to an effective 4-acceleration triggered by relative-motion effects. For the same reason, the real observers ``see'' an imperfect fluid with a non-vanishing effective energy flux (see Eq.~(\ref{lrs2}b)). On the other hand, the energy density ($\rho$), the pressure ($p$) and the viscosity ($\pi_{ab}$) of the matter remain unaffected by the frame change. In other words, $\tilde{\rho}=\rho$, $\tilde{p}=p=0$ and $\tilde{\pi}_{ab}=\pi_{ab}=0$ to linear order, with the last two equalities holding for dust~\cite{M}. We finally note that neither the Hubble nor the tilted frame need to be irrotational or shear-free (at the linear level), since none of these variables appears in our calculations.

\section{The peculiar deceleration parameter}\label{sPDP}
The differences in the volume-expansion scalars and in their time-derivatives seen in Eqs.~(\ref{Thetas}a) and (\ref{Thetas}b), imply that the expansion rates and the deceleration/acceleration rates, as measured in the (reference) Hubble frame and in that of the real observers, should differ as well.

\subsection{Two deceleration parameters}\label{ssTDPs}
The deceleration/acceleration of the expansion is monitored by the deceleration parameter. Written in the coordinate system of the smooth Hubble flow and in the rest-frame of the bulk peculiar motion, these are given by
\begin{equation}
q= -\left(1+{3\dot{\Theta}\over\Theta^2}\right) \hspace{7.5mm} {\rm and} \hspace{7.5mm} \tilde{q}= -\left(1+{3\tilde{\Theta}^{\prime}\over\tilde{\Theta}^2}\right)\,,  \label{qs}
\end{equation}
respectively (e.g.~see~\cite{TCM,EMM}). On using the linear expressions (\ref{Thetas}a) and (\ref{Thetas}b), the above implies that $\tilde{q}\neq q$. Indeed, to begin with, let us recall that $\Theta=3H$ in the  background and that $\dot{H}= -H^2(1+\Omega/2)<0$ in a Friedmann universe with dust, where $\Omega=\rho/3H^2$ represents the corresponding density parameter~\cite{TCM,EMM}. Then, keeping in mind that $|\tilde{\vartheta}|/H\ll1$ throughout the linear regime, Eqs.~(\ref{Thetas}) and (\ref{qs}) combine to give the following linear relation
\begin{equation}
\tilde{q}= q+ {\tilde{\vartheta}^{\prime}\over3\dot{H}} \left(1+{1\over2}\,\Omega\right)= q- {\tilde{\vartheta}^{\prime}\over3H^2}\,,  \label{tq2}
\end{equation}
between $\tilde{q}$ and $q$~\cite{T1,T2}. The last term in the above is essentially a ``correction term'' induced purely by relative-motion effects, which also depends on the background density parameter ($\Omega$). Recall that the ratio $|\tilde{\vartheta}^{\prime}/\dot{H}|$ is not necessarily small, even at the linear level (see footnote~4).

\subsection{Correction due to relative motion}\label{ssCRM}
In order to analyse the linear relation (\ref{tq2}) further, we need to know the time evolution of the peculiar volume scalar ($\tilde{\vartheta}$) relative to the bulk-flow frame. For zero cosmological constant and pressureless matter, the latter is given by the first-order relation~\cite{ET,TK1},
\begin{equation}
\tilde{\vartheta}^{\prime}= -H\tilde{\vartheta}+ \tilde{\rm D}^a\tilde{v}_a^{\prime}\,.  \label{lpRay1}
\end{equation}
Since the bulk-flow surveys provide the mean peculiar velocity of the motion, but not its time derivative, the second term on the right-hand side of (\ref{lpRay1}) requires additional theoretical work. We therefore turn to relativistic cosmological perturbation theory and relate the peculiar velocity field to density perturbations. In particular, linearising Eq.~(2.3.1) of~\cite{TCM} (or Eq.~(10.101) of~\cite{EMM}), in the bulk-flow frame, gives~\cite{TK2}-\cite{FT}
\begin{equation}
\tilde{\Delta}_a^{\prime}= -\tilde{\mathcal{Z}}_a -3aH\tilde{v}_a^{\prime}- 3aH^2\tilde{v}_a+ a\tilde{\rm D}_a\tilde{\vartheta}\,,  \label{laux1}
\end{equation}
given that $\tilde{q}_a=-\rho\tilde{v}_a$ to linear order (see Eq.~(\ref{lrs2}b) in \S~\ref{ssLRBTFs}) and  $\dot{\rho}=-3H\rho$ in the pressureless Friedmann background. Also, the spatial gradients $\tilde{\Delta}_a=(a/\rho)\tilde{\rm D}_a\tilde{\rho}$ and $\tilde{\mathcal{Z}}_a=a\tilde{\rm D}_a\tilde{\Theta}$ describe linear inhomogeneities in the matter density and in the universal expansion respectively, both measured in the bulk-flow frame~\cite{TCM,EMM}. Linearising the 3-divergence of the above and solving for $\tilde{\rm D}^a\tilde{v}_a^{\prime}$, we obtain~\cite{TK2}
\begin{equation}
\tilde{\rm D}^a\tilde{v}_a^{\prime}= -H\tilde{\vartheta}+ {1\over3H}\,\tilde{\rm D}^2\tilde{\vartheta}- {1\over3a^2H}\left(\tilde{\Delta}^{\prime} +\tilde{\mathcal{Z}}\right)\,,  \label{laux2}
\end{equation}
where $\tilde{\Delta}=a\tilde{\rm D}^a\tilde{\Delta}_a$ (with $a\tilde{\rm D}^a\tilde{\Delta}_a^{\prime}=\tilde{\Delta}^{\prime}$ to first order) and $\tilde{\mathcal{Z}}=a\tilde{\rm D}^a\tilde{\mathcal{Z}}_a$. In addition, $\tilde{{\rm D}}^2=\tilde{\rm D}^a\tilde{\rm D}_a$ is the 3-dimensional covariant Laplacian and $a=a(t)$ is the (background) cosmological scale factor The scalar $\tilde{\Delta}$ monitors scalar perturbations in the matter distribution, namely overdensities or underdensities, as ``seen'' by the real (i.e.~the bulk-flow) observers, while $\tilde{\mathcal{Z}}$ does the same for perturbations in the volume expansion.\footnote{Recasting  expressions (\ref{laux1}) and (\ref{laux2}) relative to the Hubble frame of  the fictitious observers, one recovers the familiar linear relation $\dot{\Delta}_a=-\mathcal{Z}_a$ and $\dot{\Delta}=-\mathcal{Z}$ (e.g.~see~\cite{TCM,EMM}), without any relative-motion terms (as expected).}

At this point, we need to emphasise that the relativistic expressions (\ref{laux1}) and (\ref{laux2}) cannot be reproduced in a Newtonian study. The reason is the different way the two theories treat the gravitational field. Newtonian gravity appeals to the gravitational potential, which couples to the matter via the Poisson equation. There is no potential in general relativity, but spacetime curvature. Also, Poisson's formula is replaced by the Einstein equations, coupling the curvature to the energy-momentum tensor of the matter. The latter also carries the contribution of the energy flux vector, which is our case is entirely due to the bulk flow (i.e.~$\tilde{q}_a= -\rho\tilde{v}_a$)). There is no flux contribution to the Newtonian gravitational field. This extra input to Einstein's equations feeds into the conservation laws and eventually into the relativistic formulae governing the evolution of peculiar velocity perturbations (see Appendix~\ref{app1} here for a comparison to the Newtonian study and also~\cite{TT,FT} for further discussion).

Overall, Eqs.~(\ref{laux1}) and (\ref{laux2}), as well as the implications resulting from them, are purely general relativistic corrections with no close Newtonian analogues. More specifically, the Newtonian propagation formula of the density gradients reads $\tilde{\Delta}^{\prime}_{\alpha}=-\tilde{\mathcal{Z}}_{\alpha}$, where $\tilde{\Delta}_{\alpha}=(a/\rho)\partial_{\alpha}\rho$ and $\tilde{\mathcal{Z}}_{\alpha}=a\partial_{\alpha}\tilde{\Theta}$, with no flux terms (equivalently no peculiar-velocity terms) on the right-hand side (e.g.~see~\cite{E} and also Appendix~\ref{app1} here). In other words, Eq.~(\ref{laux1}) lies beyond the limits of Newtonian theory.

The reader is also referred to~\cite{TKA} for a detailed comparison between the relativistic and the Newtonian analysis of the relative-motion effects on the deceleration parameter measured in the bulk-flow frame. That work explains in more detail why the two theories lead to different results and conclusions.

\subsection{Scale-dependent correction}\label{ssS-DC}
Expression (\ref{laux2}) and the 3-divergence of $\tilde{v}_a^{\prime}$, namely $\tilde{\rm D}^a\tilde{v}_a^{\prime}$, monitor the time-evolution of $\tilde{\vartheta}$ (see Eq.~(\ref{lpRay1}) above). The latter determines the dimensionless ratio $\tilde{\vartheta}^{\prime}/3\dot{H}$ in the correction term on the right-hand side of (\ref{tq2}) and ultimately decides the relative-motion effect on the deceleration parameter (see~\cite{TK2} for details). Before doing so, however, it helps to analyse the scale-dependence of the aforementioned ratio.

The Laplacian seen in Eq.~(\ref{laux2}) introduces a characteristic scale dependence, which becomes explicit after a simple Furrier decomposition. Indeed, substituting (\ref{laux2}) into (\ref{lpRay1}) and splitting the perturbed variables harmonically leads to\footnote{The perturbed variables split as $\tilde{\vartheta}=\sum_n\tilde{\vartheta}_{(n)} \mathcal{Q}^{(n)}$, $\tilde{\Delta}=\sum_n\tilde{\Delta}_{(n)} \mathcal{Q}^{(n)}$ and $\tilde{\mathcal{Z}}= \sum_n\tilde{\mathcal{Z}}_{(n)} \mathcal{Q}^{(n)}$, where $\tilde{\rm D}_a\tilde{\vartheta}_{(n)}=0= \tilde{\rm D}_a\tilde{\Delta}_{(n)}=\tilde{\rm D}_a\tilde{\mathcal{Z}}_{(n)}$. Also, $\mathcal{Q}^{(n)}$ are scalar harmonics with $\mathcal{Q}^{\prime\,(n)}=0$ and $\tilde{\rm D}^2\mathcal{Q}^{(n)}=-(n/a)^2\mathcal{Q}^{(n)}$~\cite{TCM,EMM}.}
\begin{equation}
{\tilde{\vartheta}_{(n)}^{\prime}\over3\dot{H}}= \left(1+{1\over2}\,\Omega\right)^{-1} \left\{{2\over3} \left[1+{1\over6}\left({\lambda_H\over\lambda_n}\right)^2\right] {\tilde{\vartheta}_{(n)}\over H}+ {1\over9}\left({\lambda_H\over\lambda_K}\right)^2
\left({\tilde{\Delta}_{(n)}^{\prime}\over H}
+{\tilde{\mathcal{Z}}_{(n)}\over H}\right)\right\}\,,  \label{lpRay2}
\end{equation}
for the $n$-th harmonic~\cite{TK2}. Here, $\lambda_H=3/\Theta=1/H$ is the Hubble horizon, $\lambda_n=a/n$ is the physical size of the peculiar-velocity perturbation and $\lambda_K=a/|K|$ (with $K=\pm1$) is the curvature scale of the FRW background. When the latter is very close to spatial flatness, namely for $\Omega\rightarrow1$ and $\lambda_H/\lambda_K\rightarrow0$, we may neglect the last term on the right-hand side of (\ref{lpRay2}). Then, the latter reduces to
\begin{equation}
{\tilde{\vartheta}_{(n)}^{\prime}\over\dot{H}}= {4\over3}\,\left[1+{1\over6} \left(\lambda_H\over\lambda_n\right)^2\right] {\tilde{\vartheta}_{(n)}\over H}\,,  \label{lpRay3}
\end{equation}
with $H>0$ and $\dot{H}<0$ always. The scale dependence of the above means that the dimensionless ratio $|\tilde{\vartheta}_{(n)}^{\prime}/\dot{H}|$ can exceed unity when $\lambda_n\ll\lambda_H$, despite the fact that $|\tilde{\vartheta}_{(n)}|/H\ll1$ at the linear level (see also footnote~4). This happens because Eq.~(\ref{lpRay3}) is more sensitive to the scale-ratio ($\lambda_H/\lambda_{n}$), rather than to the kinematic ratio ($\tilde{\vartheta}_{(n)}/H$). As a result, on relatively small wavelengths, the linear kinematics are dominated by relative-motion effects. An analogous effect is also observed in studies of density perturbations and leads to the familiar Jeans length (e.g.~see~\cite{TCM,EMM}). In that case, on relatively small scales (i.e.~smaller than the associated Jeans length), the pressure gradients dominate over the background gravitational pull and thus dictate the linear evolution of the density perturbations. More specifically, the pressure support halts the collapse and forces the inhomogeneities to decaying oscillations. In our case, it is the peculiar-velocity perturbations that dominate over the background Hubble expansion on small enough scales. What is most important is that the scale-dependence seen in (\ref{lpRay3}) readily transfers into Eq.~(\ref{tq2}), which now reads
\begin{equation}
\tilde{q}_{(n)}= q+ {2\over3} \left[1+{1\over6}\left(\lambda_H\over\lambda_n\right)^2\right] {\tilde{\vartheta}_{(n)}\over H}\,.  \label{tq3}
\end{equation}
This formula provides the deceleration parameter measured by observers inside a large-scale bulk flow in a nearly-flat, almost-FRW universe filled with ordinary dust. Given that $\tilde{\vartheta}_{(n)}/H\ll1$ always, on lengths close and beyond the Hubble horizon (i.e.~for $\lambda_H/\lambda_n\lesssim1$) the relative-motion effects fade away and the correction term on the right-hand side of (\ref{tq3}) becomes completely negligible. Therefore, on super-Hubble lengths $\tilde{q}_{(n)}\rightarrow q$, as expected. On the other hand, the correction term in Eq.~(\ref{tq3}) can lead to a considerable difference between the local ($\tilde{q}_{(n)}$) and the global ($q$) deceleration parameter on wavelengths well inside the Hubble horizon. Indeed, on subhorizon scales expression (\ref{tq3}) reduces to
\begin{equation}
\tilde{q}_{(n)}= q+ {1\over9}\left(\lambda_H\over\lambda_n\right)^2 {\tilde{\vartheta}_{(n)}\over H}\,,  \label{tq4}
\end{equation}
where $\tilde{\vartheta}_{(n)}$ is positive/negative inside (slightly) expanding/contracting bulk flows. Clearly, when $\lambda_H/\lambda_n\gg1$, the correction term on the right-hand side of the above can become large even for $|\tilde{\vartheta}_{(n)}|/H\ll1$. This happens on scales of few hundred Mpc, for example, which are well inside the Hubble horizon, but still outside the nonlinear range (see next section and Table~\ref{tab1} there).

\section{The peculiar Jeans length}\label{sPJL}
The Jeans length is a familiar linear result, marking the scale below which the pressure gradients take over the background gravity and dictate the evolution of density perturbations. An analogous scale also emerges when dealing with peculiar-velocity perturbations.

\subsection{The critical length scale}\label{ssCLS}
Once the local expansion/contraction rate of a bulk flow is known, namely given the magnitude of $\tilde{\vartheta}_{(n)}$, one can use Eq.~(\ref{tq4}) to estimate the critical length below which the linear relative-motion effects dominate over the background expansion and can drastically change the local value of $\tilde{q}_{(n)}$. Typically, this happens when the correction term seen in (\ref{tq4}) equals (in absolute value) the deceleration parameter measured in the Hubble frame, namely when
\begin{equation}
{1\over9}\left(\lambda_H\over\lambda_n\right)^2 {|\tilde{\vartheta}_{(n)}|\over H}= q\,.  \label{lambdaq1}
\end{equation}
Assuming that the above happens at $\lambda_n=\lambda_P$, the critical length is
\begin{equation}
\lambda_P= \sqrt{{1\over9q}\,{|\tilde{\vartheta}_{(n)}|\over H}}\,\lambda_H\,,  \label{lambdaq2}
\end{equation}
keeping in mind that $q>0$ in FRW cosmologies with conventional matter and no cosmological constant. Physically, $\lambda_P$ is closely analogous to the Jeans length ($\lambda_J$), familiar from the linear study of density perturbations.\footnote{A comparison between the standard Jeans length and its peculiar counterpart is due here. The former is given by $\lambda_J=c_s\lambda_H/\sqrt{3}$, where the numerical factor depends on the equation of state of the matter~\cite{TCM,EMM}. Note that $c_s$, with $c_s^2=\dot{p}/\dot{\rho}$, is the effective sound speed. The latter is dimensionless and satisfies the constraint $c_s<1$, since we have set the velocity of light to unity. Comparing $\lambda_J$ to definition (\ref{lambdaq2}), shows that both scales are given as simple fractions of the Hubble radius. The main difference is that, in (\ref{lambdaq2}), the role of the sound speed is played by the dimensionless ratio $|\tilde{\vartheta}_{(n)}|/H$.} What distinguishes them is that $\lambda_J$ marks the scale where linear pressure gradients dominate over the background gravity, while $\lambda_P$ sets the threshold below which linear peculiar-velocity perturbations take over the background Hubble expansion. We will therefore refer to $\lambda_P$ as the \textit{peculiar Jeans length}, to underline the aforementioned  analogy between these two characteristic scales.

\subsection{The transition scale}\label{ssTS}
An additional crucial point following from Eq.~(\ref{tq4}) is that the overall effect of relative motion on the local value of $\tilde{q}_{(n)}$ also depends on the sign of $\tilde{\vartheta}_{(n)}$. In particular, when combined, relations (\ref{tq4}) and (\ref{lambdaq2}) imply that
\begin{equation}
\tilde{q}= \tilde{q}^{(+)}> 2q\,,  \label{tq+1}
\end{equation}
when
\begin{equation}
\lambda< \lambda_P \hspace{7.5mm} {\rm and} \hspace{7.5mm} \tilde{\vartheta}>0\,,  \label{tq+2}
\end{equation}
having dropped the mode-index $(n)$ for the economy of the presentation. The same relations also ensure that
\begin{equation}
\tilde{q}= \tilde{q}^{(-)}< 0\,,  \label{tq-1}
\end{equation}
when
\begin{equation}
\lambda<\lambda_P \hspace{7.5mm} {\rm and} \hspace{7.5mm} \tilde{\vartheta}<0\,.  \label{tq-2}
\end{equation}
Therefore, the value of $\tilde{q}$ can be twice as large (or even larger) as that of $q$ inside (slightly) expanding bulk flows (i.e.~those with $\tilde{\vartheta}>0$).\footnote{Hereafter $\tilde{q}^{(+)}$ will denote the deceleration parameter measured inside an expanding bulk flow (with $\tilde{\vartheta}>0$), while $\tilde{q}^{(-)}$ will correspond to contracting peculiar motions (with $\tilde{\vartheta}<0$).} In contrast, $\tilde{q}$ can take negative values (while $q$ is still positive) in slightly contracting bulk motions (with $\tilde{\vartheta}<0$). In the latter case the peculiar Jeans length also defines the ``transition'' scale, where $\tilde{q}^{(-)}$ changes from positive to negative (see Table~\ref{tab1} and also Fig.~\ref{fig:PJeans} below).

Clearly, the possibility of a sign change for the (local) deceleration parameter is most intriguing and deserves further attention. Confining to sub-Hubble scales, Eq.~(\ref{tq4}) and definition (\ref{lambdaq2}) combine to the alternative expression
\begin{equation}
\tilde{q}^{(-)}= q\left[1-\left({\lambda_P\over\lambda}\right)^2\right]\,,  \label{tq-}
\end{equation}
for (slowly) contracting bulk flows with $\tilde{\vartheta}<0$. One can now readily see that the local deceleration parameter is always smaller than the one measured by the Hubble-flow observers (i.e.~$\tilde{q}^{(-)}<q$). Nevertheless, on scales much larger than the associated peculiar Jeans length (that is for $\lambda_P\ll\lambda\ll\lambda_H$) the difference is very small and the two parameters essentially coincide (with $\tilde{q}^{(-)}\rightarrow q$). As we move to progressively smaller distances, however, the difference between $\tilde{q}^{(-)}$ and $q$ keeps increasing, with the local deceleration parameter crossing the $\tilde{q}^{(-)}=0$ threshold at the transition scale (i.e.~at the peculiar Jeans length where $\lambda=\lambda_P$). These two features are therefore characteristic phenomenological predictions of the bulk-flow scenario.

\begin{figure}[tbp]
\centering \vspace{6.5cm} \includegraphics{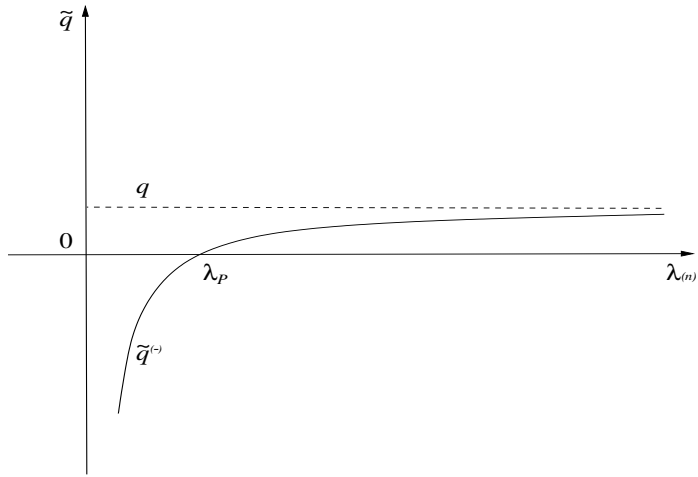} \caption{The deceleration parameter ($\tilde{q}^{(-)}$) measured in the frame of a slowly contracting bulk flow with peculiar Jeans length $\lambda_P$ (see Eq.~(\ref{tq-})). On large scales, far away from the observer, $\tilde{q}^{(-)}$ approaches the deceleration parameter of the Einstein-de Sitter background (horizontal dashed line). In other words, $\tilde{q}^{(-)}\rightarrow q=1/2$ when $\lambda\gg\lambda_P$. On the other hand, as we move closer to the observer, the local deceleration parameter starts to diverge, dropping below the $\tilde{q}^{(-)}=0$ mark and turning negative at the transition scale ($\lambda_P$). Note that we have assumed that $\tilde{\vartheta}=$~const. and normalised so that $\lambda_P=1$.}  \label{fig:qplot}
\end{figure}

Following (\ref{tq-}), deep inside $\lambda_P$ the local deceleration parameter can drop well below zero, although eventually the nonlinear effects will take over and our linear analysis will no longer apply. Typically the nonlinear scale is believed to lie below the 100~Mpc mark. Qualitatively, this behaviour agrees with our expectations that the peculiar-motion effects get stronger on smaller lengths. Assuming that $\tilde{\vartheta}$ is scale invariant, the evolution of $\tilde{q}^{(-)}$ in terms of scale/distance follows the solid curve depicted in Fig.~\ref{fig:qplot}, the shape of which reflects the scale-dependence seen in Eq.~(\ref{tq-}). The numerics depend on the specifics of the peculiar motion. These include the location of the transition scale ($\lambda_P$), which is fixed once the local expansion/contraction rate ($\tilde{\vartheta}$) is given. As we mentioned above, the latter has been treated here as a constant, since it is still very difficult to extract from the data. Nevertheless, in a complete scenario, $\tilde{\vartheta}$ would probably have a scale dependence as well.

Note that the profile depicted in Fig.~\ref{fig:qplot} qualitatively resembles those of the deceleration parameters reconstructed from the supernovae data (e.g.~see~\cite{GW}-\cite{B}). These reconstructions, however, typically introduce a two-parameter ansatz for the $q$-distribution. Here, the $\tilde{q}^{(-)}$-profile follows naturally from the scale-dependence of Eq.~(\ref{tq-}).

\subsection{The local deceleration parameter}\label{ssLDP}
\begin{figure}[tbp]
\centering \vspace{6cm} \includegraphics{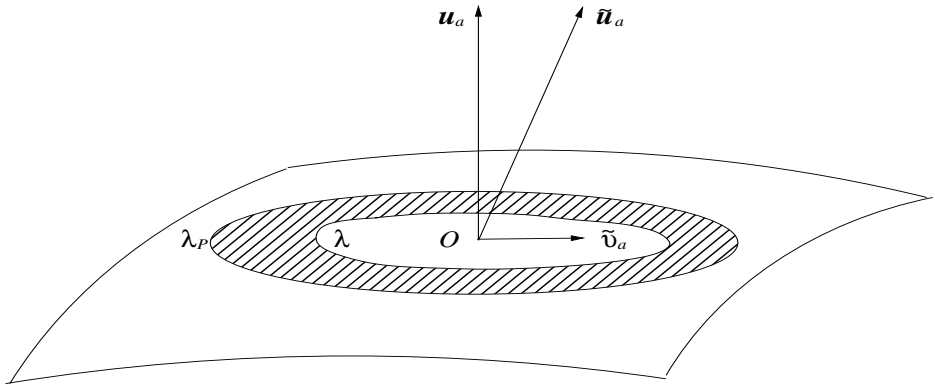} \caption{Observer ($O$) inside a bulk flow (central section) like those quoted in Table~\ref{tab1}, with 4-velocity $\tilde{u}_a$ and peculiar velocity $\tilde{v}_a$, relative to the smooth Hubble expansion (identified with the $u_a$-field). Around the observer there is a spherical region (shaded area), with size determined by the associated peculiar Jeans length ($\lambda_P$) and corresponding to redshift $z_{\lambda_P}$. In the case of contracting bulk flows, the outer limits of the shaded domain mark the transition scale, where the locally measured deceleration parameter ($\tilde{q}^{(-)}$) turns from positive to negative (see Table~\ref{tab1}). Put another way, the value of $\tilde{q}^{(-)}$ becomes progressively less negative away from the observer at $O$, it crosses the zero threshold at the transition scale and starts taking positive values beyond $\lambda_P$, eventually approaching $q=1/2$ on large enough lengths (see Fig.~\ref{fig:qplot}). The unsuspecting observer may then be mislead to believe that the universe started to accelerate at $z=z_{\lambda_P}$ and in so doing misinterpret the local change in the sign of $\tilde{q}^{(-)}$ as recent global acceleration.}   \label{fig:PJeans}
\end{figure}

Bulk-flow surveys provide the mean peculiar velocity but not its divergence. There are systematic uncertainties with these higher moments because of the noisy peculiar-flow field~\cite{FWH}. Here, we will approximate $\tilde{\vartheta}$ from the measured mean peculiar velocity by means of typical dimensional-analysis arguments. In particular, given that the spatial curvature is negligible well inside the Hubble horizon, we will set $\tilde{\vartheta}\simeq \partial_{\alpha}\tilde{v}^{\alpha}\simeq \pm\sqrt{3}\langle\tilde{v}\rangle/\lambda$.\footnote{Ignoring spatial curvature gives $\tilde{\vartheta}\simeq \partial_x\tilde{v}^x+\partial_y\tilde{v}^y+\partial_z\tilde{v}^z$. Assuming that $\tilde{v}^x\simeq\tilde{v}^y\simeq \tilde{v}^z\simeq\tilde{v}$, we may set $\tilde{\vartheta}\simeq\pm3\tilde{v}/\lambda$ and $\langle\tilde{v}\rangle\simeq\sqrt{3}\tilde{v}$, which combine to give $\tilde{\vartheta}\simeq \pm\sqrt{3}\langle\tilde{v}\rangle/\lambda$. Then, for a mean bulk-flow velocity of, say, 240~km/sec on scales of 200~Mpc (see~\cite{Setal} and also Table~\ref{tab1}), we obtain $\tilde{\vartheta}\simeq\pm2.1$~km/secMpc.} Note that $\langle\tilde{v}\rangle$ is the mean peculiar velocity and the plus/minus sign indicates expanding/contracting bulk flows. Then, since $v_H=\lambda H$ is the Hubble velocity on scale $\lambda$, Eqs.~(\ref{tq4}) and (\ref{lambdaq2}) become
\begin{equation}
\tilde{q}^{(\pm)}\simeq q\pm {\sqrt{3}\over9}\left({\lambda_H\over\lambda}\right)^2 {\langle\tilde{v}\rangle\over v_H}  \label{tq5}
\end{equation}
and
\begin{equation}
\lambda_P\simeq \sqrt{{\sqrt{3}\over9q} {\langle\tilde{v}\rangle\over v_H}}\,\lambda_H\,,  \label{plambdaJ}
\end{equation}
respectively. We can now turn to the observations to estimate the approximate values of $\tilde{q}^{(\pm)}$ and $\lambda_P$, based on some of the reported bulk-flow measurements (see Table~\ref{tab1}). Typically, these claim velocities between 240 and 410~km/sec over regions varying from 140 to 280~Mpc in their diameter (e.g.~see~\cite{FWH}-\cite{Setal}). Assuming slightly expanding bulk flows, we find that the deceleration parameters measured by observers located at the centre of these bulk flows lie in the range $+1.01\lesssim \tilde{q}^{(+)}\lesssim+7.08$ (3rd column in Table~\ref{tab1}). For slightly contracting peculiar motions, on the other hand, Eq.~(\ref{tq5}) assigns negative values to the local deceleration parameter, with $-6.08\lesssim \tilde{q}^{(-)}\lesssim-0.01$ (4th column in Table~\ref{tab1}). In the latter case, the peculiar Jeans length (see definition (\ref{plambdaJ}) also marks the corresponding transition scales (where the deceleration parameter crosses the $\tilde{q}=0$ divide). These vary between 282 and 508~Mpc (5th column in Table~\ref{tab1}).\footnote{Although in all the examples quoted in Table~\ref{tab1} the transition scale exceeds that of the reported bulk-flow measurements (see also Fig.~\ref{fig:PJeans}), it does not necessarily need to be so.} Note that, in all cases, the universe is assumed to decelerate globally, with $q=+0.5$ in the reference $u_a$-frame. Therefore, the over-deceleration, or the acceleration, seen in Table~\ref{tab1} are not real but local artifacts of the observers' relative motion. Nevertheless, the affected scales are large enough ($\lambda_P$ is of the order of few to several hundred Mpc -- perhaps even larger due to possible projection effects) to create the false impression that there was a recent global change in the expansion rate of the whole universe.

\begin{table}
\caption{Representative estimates of the deceleration parameter ($\tilde{q}$ -- see Eq.~(\ref{tq5})) measured in the rest-frame of the bulk flows reported in~\cite{FWH}-\cite{Setal}, with $\tilde{q}^{(+)}$ corresponding to slightly expanding and $\tilde{q}^{(-)}$ to slightly contracting bulk flows. In the latter case the peculiar Jeans length ($\lambda_P$ -- see last column) also marks the transition scale, where the sign of $\tilde{q}^{(-)}$ turns negative (see Eq.~(\ref{plambdaJ}) and also Fig.~\ref{fig:PJeans}). Note that both $\lambda$ and $\lambda_P$ are measured in Mpc, while the mean bulk velocity ($\langle\tilde{v}\rangle$) is given in km/sec. Also, in all cases, the background universe is assumed to decelerate with $q=+0.5$. Finally, we have set $H\simeq70$~km/sec\,Mpc and $\lambda_H=1/H\simeq4\times10^3$~Mpc today.}
\begin{center}\begin{tabular}{cccccccc}
\hline \hline & \hspace{-20pt} Survey & $\lambda$ & $\langle\tilde{v}\rangle$ & $\tilde{q}^{(+)}$ & $\tilde{q}^{(-)}$ & $\lambda_P$ &\\ \hline \hline & $\begin{array}{c} \hspace{-20pt} {\rm Nusser\,\&\,Davis} \\ \hspace{-20pt} {\rm Colin,\,et\,al} \\ \hspace{-20pt} {\rm Scrimgeour,\,et\,al} \\ \hspace{-20pt} {\rm Ma\,\&\,Pan} \\ \hspace{-20pt} {\rm Turnbull,\,et\,al} \\ \hspace{-20pt} {\rm Feldman,\,et\,al} \end{array}$ & $\begin{array}{c} 280 \\ 250 \\ 200 \\ 170 \\ 140 \\ 140 \end{array}$ & $\begin{array}{c} 260 \\ 260 \\ 240 \\ 290 \\ 250 \\ 410 \end{array}$ & $\begin{array}{c} +1.01 \\ +1.24 \\ +1.81 \\ +3.05 \\ +4.58 \\ +7.08 \end{array}$ & $\begin{array}{c} -0.01 \\ -0.24 \\ -0.81 \\ -2.05 \\ -3.58 \\ -6.08 \end{array}$ & $\begin{array}{c} 282 \\ 304 \\ 323 \\ 384 \\ 400 \\ 508 \end{array}$\\ [2.5truemm] \hline \hline
\end{tabular}\end{center}\label{tab1}\vspace{-10pt}
\end{table}

Following (\ref{tq4}) and (\ref{tq5}), the numerical estimates of $\tilde{q}$ seen in Table~\ref{tab1} depend on the dimensionless ratios $\tilde{\vartheta}/H$ and $(\lambda_H/\lambda)^2$, which makes them more sensitive to the latter rather than the former.\footnote{In Table~\ref{tab1} the surveys are cited in order of descending bulk-flow scale, which shows the impact of the peculiar motion to drop with increasing scale. More specifically, the weakest effect corresponds to the survey of Nusser \& Davis~\cite{ND}. There, the deceleration parameter becomes only marginally negative ($\tilde{q}^{(-)}\simeq-0.01$) and the transition scale just exceeds that of the reported bulk flow ($\lambda_p\simeq282~{\rm Mpc}\gtrsim\lambda\simeq280~{\rm Mpc}$). At the opposite end lies the survey of Feldmann~et~al~\cite{FWH}, where the impact of the relative-motion on $\tilde{q}$ and $\lambda_P$ is very strong.} According to Eqs.~(\ref{lambdaq2}) and (\ref{plambdaJ}), on the other hand, the estimates for $\lambda_P$ seen in Table~\ref{tab1} are sensitive to the $\tilde{\vartheta}/H$ ratio only. Given the values of $\lambda$ and $\tilde{\vartheta}$, both of these ratios are determined by the magnitude of the Hubble parameter. Qualitatively speaking, the smaller the value of $H$, the larger those of $\lambda_H/\lambda$ and $\tilde{\vartheta}/v_H$ and therefore the stronger the relative-motion effect on the local deceleration parameter ($\tilde{q}^{(\pm)}$) and the larger the peculiar Jeans length ($\lambda_P$). If the Hubble parameter was to increase, the situation will be reversed and the bulk-flow impact will weaken.\footnote{The estimated values of $\tilde{\vartheta}$ have been obtained from the measured mean bulk velocity, using dimensional-analysis arguments (see footnote~10). The numbers for $\tilde{\vartheta}/H$ used in Table~\ref{tab1} may therefore change when direct measurements of $\tilde{\vartheta}$ become available. Here, to compensate for a possible overestimation of $\tilde{\vartheta}$, we have identified $\lambda$ with the diameter rather then the radius of the reported bulk-flow measurements, which reduces the overall relative-motion effects seen in Table~\ref{tab1}. Alternatively, one can account for the present ambiguity by introducing an extra parameter ($\alpha$). More specifically, we may write $\tilde{\vartheta}\simeq \pm\alpha\sqrt{3}\langle\tilde{v}\rangle/\lambda$, with $0<\alpha<1$ to avoid overestimating the impact of the peculiar flow on the local deceleration parameter (e.g.~see~\cite{TK2}). Then, setting $\alpha=1/2$ reduces the impact on $\tilde{q}$ by half, assuming that $\alpha=1/4$ drops the effect by another half and so on (see also~\cite{TK2}).}

Before closing this section, we should remind the reader that the negative values of the deceleration parameter, quoted in the 4th column of Table~\ref{tab1}, have been obtained within a conventional almost-FRW cosmology. There was no need to violate the strong-energy condition (we have assumed dust), to introduce a cosmological constant, or to modify general relativity in any way. The inferred accelerated expansion is not real, since the universe is still globally decelerating, but a local artifact of the observers' peculiar motion relative to the smooth universal (Hubble) flow. This scenario was originally proposed in~\cite{T1,T2} and it was later refined in~\cite{TK2}. Recently, an alternative approach, involving the null geodesics of electromagnetic signals (emitted and received by observers in the universe) and introducing a multiple-moment expansion of the corresponding luminosity distance, arrived at analogous results~\cite{H}. In particular, the study showed that the (effective) deceleration parameter associated with the aforementioned null rays could take negative values within an otherwise decelerating universe.

\subsection{The signatures of the bulk-flow scenario}\label{ssSB-FS}
One might ask whether there is a way for the unaware bulk-flow observers to find out that they have been merely experiencing an illusion. The answer should be in the data, which should contain the ``signatures'' of relative motion. A first indication could come from the observed distribution of the deceleration parameter in terms of scale. Qualitatively speaking, the deceleration parameter should get progressively more negative on smaller scales closer to the observer. In more quantitative terms, the scale-distribution of the deceleration parameter should resemble the profile of the solid curve depicted in Fig.~\ref{fig:qplot}. The shape of the latter, which corresponds to the simplest case where $\tilde{\vartheta}=$~const., resembles the redshift-space distribution of the deceleration parameters reconstructed from the supernovae data (e.g.~see~\cite{GW}-\cite{B} and references therein). There, the reconstructions were achieved by introducing an ansatz for the description of $q$, typically in the form of a two-parameter function. Here, the profile of the solid curve depicted in Fig.~\ref{fig:qplot} is a natural result of the peculiar-motion effects.

A second signature of the bulk-flow scenario, which should also be sought in the data, is an apparent (Doppler-like) dipolar anisotropy in the sky-distribution of the deceleration parameter. More specifically, in the data, the value of $\tilde{q}$ should be smaller than the average towards a certain point in the sky and equally larger towards the antipodal. Put another way, the universe should appear to accelerate faster in one direction and equally slower in the opposite. This happens because observers inside the bulk flow have their own individual peculiar velocities, which are generally different (both in magnitude and direction) from the mean bulk velocity.\footnote{Strictly speaking, only observers located at the centre of a smooth spherical peculiar flow and moving with the mean velocity of the bulk will see no anisotropies in their local distribution of the deceleration parameter ($\tilde{q}$). In any other case, the observer should be able to detect some degree of anisotropy for a number of reasons. For instance, the measured $\tilde{q}$-distribution may differ from one hemisphere to the other because the observer is not at the centre of the bulk-flow domain, or because the latter is not spherically symmetric. Also, there could be shear-like anisotropy because the peculiar motion itself is generically anisotropic. Of all the possible types of anisotropy, however, the dipolar one is the characteristic ``trademark'' signature of relative motion.} For typical bulk-flow observers and on large enough scales, the aforementioned peculiar-velocity difference should be small, thus leading to a weak dipole-like anisotropy in the observed distribution of the deceleration parameter~\cite{T2}. Moreover, assuming that the Cosmic Microwave Background (CMB) dipole seen by these observers is also an apparent relative-motion effect, the two dipole axes should not lie far apart. When dealing with atypical bulk-flow observers, however, with individual peculiar velocities considerably different than the mean (in their magnitude or/and in their direction), the apparent dipolar anisotropy could be large. The possibility that we might be such atypical observers cannot be excluded~\cite{Seetal}.

In retrospect, the presence of an apparent (Doppler-like) dipole in the sky-distribution of the deceleration parameter due to the observer's peculiar flow, is intuitively plausible. After all, relative motions are typically associated with (apparent) dipolar anisotropies. The emergence of such an anisotropy was first suggested in~\cite{T1} and then theoretically demonstrated in~\cite{T2}, by taking into account apparent shear-like effects of the relative motion. More recently, an alternative (also fully relativistic) approach employed the null geodesics of the photon signals to analyse the impact of the peculiar-velocity field on the luminosity distance~\cite{H}. This study independently confirmed the theoretical prediction of a dipolar anisotropy in the sky-distribution of the deceleration parameter, triggered solely by the observer's relative motion.

In addition to the aforementioned theoretical predictions~\cite{T1,T2,H}, there have been reports claiming that asymmetries and a dipolar axis (fairly close to that of the CMB) may exist in the supernovae data~\cite{SW}-\cite{BBA}. In other words, our universe may indeed seem to accelerate faster towards one direction and equally slower in the opposite. This claim was specifically related to peculiar velocity fields recently in~\cite{CMRS1}. There, after re-examining the Joint Lightcurve Analysis (JLA) data of type Ia supernovae, the authors detected a fairly strong dipole in the distribution of the deceleration parameter that was closely aligned with that of the CMB. At the same time, the statistical significance of the $q$-monopole dropped, thus increasing the possibility the inferred universal acceleration to be an artefact of our peculiar motion relative to the smooth Hubble flow. Future observations, providing more and better data, should help clarify the $q$-dipole debate, which is still open~\cite{RH,CMRS2}.

As a closing comment, we would like to point out that relative-motion effects can also induce an apparent (Doppler-like) dipolar anisotropy in the sky-distribution of the Hubble parameter. The principle and the mechanism are the same with those leading to the $\tilde{q}$-dipole discussed here. After all, the two parameters are directly related, so the presence of a dipole axis in one should almost unavoidably imply a dipolar anisotropy in the sky-distribution of the other. Then, observers inside a large-scale bulk flow may also see their universe expanding faster along a given direction in the sky and equally slower in the opposite, entirely due to their peculiar motion. Indications that such a Hubble-dipole may actually exist in the data were recently reported in~\cite{Metal1,Metal2}. Moreover, the observed anisotropy is consistent with a bulk flow of several hundred km/sec extending out to several hundred Mpc~\cite{Metal2}.

\section{Discussion}\label{sD}
Bulk peculiar flows are commonplace in our universe. Such coherent large-scale motions are theoretically predicted as the inevitable byproduct of the ongoing process of structure formation and they have been repeatedly reported by a large number of surveys. The latter seem to agree on the direction of these motions, but not on their scale and on the magnitude of the associated velocities (e.g.~see~\cite{FWH}-\cite{LAH} for a representative though incomplete list).

Relative motions have long been known to introduce apparent effects that an unsuspecting observer may misinterpret as ``reality''. The history of astronomy is full of such examples. Then, given that the reported bulk motions extend out to several hundreds of Mpc, it is conceivable that living inside one of them could have ``contaminated'' our cosmological data. For instance, it is straightforward to show that the expansion and the deceleration/acceleratiion rates, measured locally by observers located inside the bulk-flow domains, are generally different from those of the actual universe, due to relative-motion effects alone (see Eqs.~(\ref{Thetas}a) and (\ref{Thetas}b) in \S~\ref{ssLRBTFs}). The question is whether (and under what circumstances) such apparent differences can become large enough to cause a serious misinterpretation of the incoming data.

With these in mind, we employed linear relativistic cosmological perturbation theory to study the effects of large-scale peculiar motions on the deceleration parameter of the universe, as measured by observers living inside these bulk flows. We found that the peculiar motion introduces a characteristic scale (the peculiar Jeans length ($\lambda_P$) -- see Eq.~(\ref{lambdaq2})), below which the local kinematics are dominated by relative-motion effects. The size of the aforementioned  critical scale, which is closely analogous to the familiar Jeans length, is determined by the velocity of the drifting domain. Using data reported by recent bulk-flow surveys, we have estimated the peculiar Jeans length ($\lambda_P$) to vary between few and several hundred Mpc (see Table~\ref{tab1} above), while it is conceivable that projection effects could extend its range even further. On scales smaller than $\lambda_P$ the interpretation of the cosmological data, by observers living inside the bulk flow, can be drastically contaminated by relative-motion effects. More specifically, the value of the deceleration parameter measured in a slightly expanding bulk flow can be significantly larger than the one measured in the frame of the Hubble expansion. Inside contracting bulk motions, on the other hand, the effect is reversed. There, the deceleration parameter drops well below the value of its Hubble-frame counterpart and it can even become negative (inside the associated peculiar Jeans length -- see Figs.~\ref{fig:qplot} and~\ref{fig:PJeans}).\footnote{Observers inside slightly expanding peculiar flows may misinterpret their overall faster expansion rate as over-deceleration of the surrounding universe. Those living in slightly contracting bulk motions, on the other hand, may misinterpret their slower local expansion as global under-deceleration (or even acceleration) of the background cosmos. Intuitively speaking, one could think of these bulk-flow observers as passengers sitting at the back of a car driving down a motorway. Then, if the speed of their car drops below the average, the unsuspecting passengers could be mislead to believe that the rest of the vehicles are accelerating away (and vice versa).}

The overall strength of the effect, which is purely relativistic with no Newtonian analogue (see Appendix~\ref{app2} here for a comparison of the two treatments and also~\cite{TKA} for a more detailed discussion) depends on the speed of the bulk flow. Given that peculiar velocities drop with increasing redshift, the impact of the relative motion grows on progressively smaller lengths closer to the observer (see Table~\ref{tab1} for representative values). More specifically, the scale distribution of the deceleration parameter, as measured by the bulk-flow observers, follows from Eq.~(\ref{tq-}) and it is depicted in Fig.~\ref{fig:qplot}. Even in the simplest case of $\tilde{\vartheta}=$~const. considered here, the profile of the solid curve in that figure resembles those of the deceleration parameters reconstructed from the supernovae data in~\cite{GW}-\cite{B}. In particular, the local deceleration parameter tends towards its background value on large enough scales and turns negative closer to the observer. This behaviour is a phenomenological prediction of the bulk-flow scenario. Another is an apparent (Doppler-like) dipole in the sky-distribution of the deceleration parameter, triggered by the direction of the observer's relative motion. Put another way, the bulk-flow observers should ``see'' their universe to accelerate faster along one direction in the sky and equally slower in the opposite. Moreover the dipolar axis should not lie far from its CMB counterpart. There have been reports in the literature that such a preferred axis may actually exist in the supernovae data~\cite{SW}-\cite{CMRS1}. Intriguingly, an analogous dipolar anisotropy, this time in the sky-distribution of the Hubble parameter, was also recently reported~\cite{Metal1,Metal2}

The attractive features of the bulk-flow scenario are that it operates within standard general relativity and within the linear regime of a perturbed Einstein-de Sitter cosmology. There is no need for new physics, or for appealing to exotic forms of matter. The inferred acceleration is not global, but a local artifact of the observers peculiar motion. As a result, there is no ``coincidence problem'' either, since the transition from deceleration to (apparent) acceleration occurs naturally at the peculiar Jeans length ($\lambda_p$). Moreover, the profile of the predicted scale-distribution of the deceleration parameter appears to agree (at least qualitatively) with those reconstructed from the supernovae data. There are also  caveats however. The aforementioned transition scale appears smaller than the one typically inferred from the reconstructed data, although it is conceivable that projection effects could explain the difference. Also, the numerics depend on the divergence ($\tilde{\vartheta}$) of the associated peculiar-velocity field, which is still very difficult to extract. Therefore, at this stage, the signatures of the bulk-flow scenario are the scale-distribution of the deceleration parameter and the apparent (Doppler-like) dipole in its sky-distribution.

Assuming that there is no natural bias in favour of expanding, or contracting, bulk flows on cosmologically relevant scales, namely those with $\lambda\gtrsim100$~Mpc, the chances of residing inside either of them should be close to 50\%. Then, according to the bulk-flow scenario, nearly half of the observers living in an otherwise Einstein de Sitter universe (with $q=1/2$ relative to the Hubble frame) may think that their cosmos is over-decelerated, with $\tilde{q}>1/2$ in their local coordinate system. The other half, on the other hand, will measure $\tilde{q}<1/2$ in their own rest-frame and they could be misled to believe that the universal expansion is under-decelerated. In fact, some of the latter observers may even measure $\tilde{q}<0$, in which case they could be erroneously forced to think that the whole universe has recently entered a phase of  accelerated expansion.

Overall, this work has, if nothing else, argued for the potentially pivotal implications of large-scale peculiar motions for the kinematics of our universe. Relativistic structure formation studies (either covariant or metric-based) are abundant in the literature. At the linear perturbative level, these treatments are also known to agree with each other (e.g.~see~\cite{BDE} for a comparison). Nevertheless, to the best of our knowledge, relatively few studies focus on the implications of the bulk peculiar flows and the reasons vary. For example, some treatments are performed in the so-called comoving gauge, where peculiar velocities are zero by default (e.g.~see~\cite{RV}). Others, although allowing for multi-fluid systems, are done in the Landau-Lifshitz (or energy) frame, where the total flux of the species vanishes (e.g.~see~\cite{TCM,EMM} and also~\cite{CMV}). There are also treatments that introduce an approximation, in the form of an effective gravitational potential analogous to that of the Newtonian studies, to account for the effects of peculiar motions (e.g.~see~\cite{M,ET}). In all these cases, the role of the bulk peculiar flows is either bypassed or downgraded and, as a result, their full implications remain largely unaccounted for.\vspace{5mm}

\noindent\textbf{Acknowledgments:} The author wishes to thank Roy Maartens, Roya Mohayaee, Leandros Perivolaropoulos, Mohamed Rameez and Subir Sarkar for helpful discussions and comments. This work was supported by the Hellenic Foundation for Research and Innovation (H.F.R.I.), under the ``First Call for H.F.R.I. Research Projects to support Faculty members and Researchers and the procurement of high-cost research equipment grant'' (Project Number: 789).

\appendix

\section{Comparing to the Newtonian study}\label{app1}
In relativity, gravity is not a force but the manifestation of spacetime curvature. The latter couples to matter via its energy-momentum tensor and Einstein's equations. Newtonian gravity, on the other hand, is a force triggered by gradients in the gravitational potential, which couples to matter through Poisson's formula. Moreover, only the density of the matter contributes to the Newtonian gravitational field, whereas in relativity there is additional input form the pressure (both isotropic and anisotropic) and from the energy flux. The latter contribution must be accounted for, especially when dealing with bulk peculiar flows, since then there is nonzero energy flux by default.

When looking into the effects of relative motion on the deceleration parameter, the Newtonian study proceeds in close parallel with the relativistic up to a certain point. This reflects the fact that, leaving the differences in the definitions aside, the Galilean transformation $\tilde{u}_{\alpha}=u_{\alpha}+\tilde{v}_{\alpha}$ is formally identical with the linearised Lorentz boost (see Eq.~(\ref{4vels}) in \S~\ref{ssRMOs}). As a result, and given that the definitions (\ref{qs}) of the two deceleration parameters hold in the Newtonian analysis as well, one arrives at the linear relations~\cite{TKA}
\begin{equation}
\tilde{q}= q- {\tilde{\vartheta}^{\prime}\over3H^2}\,,  \label{appA1}
\end{equation}
and
\begin{equation}
\tilde{\vartheta}^{\prime}= -H\tilde{\vartheta}+ \partial^{\alpha}\tilde{v}_{\alpha}^{\prime}\,,  \label{appA2}
\end{equation}
with the primes denoting convective derivatives along the $\tilde{u}_{\alpha}$ field (e.g.~$\tilde{\vartheta}^{\prime}= \partial_t\tilde{\vartheta}+ \tilde{u}^{\alpha}\partial_{\alpha}\tilde{\vartheta}$). The above are formally identical to their relativistic counterparts (compare to relations (\ref{tq2}) and (\ref{lpRay1}) respectively), provided the differences in the definitions are accounted for. Therefore, up to this point the two studies are essentially indistinguishable.

The Newtonian analysis starts to diverge form the relativistic when gravity comes into play. In particular, the absence of any flux contribution to the Newtonian gravitational field, implies that the Newtonian continuity and Euler equations differ from their relativistic analogues, even in the absence of pressure (e.g.~see~\cite{E2} as well as~\cite{TKA}). More specifically, in contrast to the relativistic conservation laws, there is no flux contribution to their Newtonian counterparts. This in turn ensures that, in Newtonian theory, the evolution of the peculiar velocity is given by
\begin{equation}
\tilde{v}_{\alpha}^{\prime}= -H\tilde{v}_{\alpha}- \partial_{\alpha}\Phi\,,  \label{appA3}
\end{equation}
where $\Phi$ is the gravitational potential. Note that this expression is identical to those obtained in typical Newtonian studies (e.g.~see~\cite{P,NDBB}), provided the latter are written in physical rather than comoving coordinates. Taking the divergence of (\ref{appA3}) and employing Poisson's formula we obtain
\begin{equation}
\partial^{\alpha}\tilde{v}_{\alpha}^{\prime}= -H\tilde{\vartheta}-{1\over2}\,\kappa\rho\delta\,,  \label{appA4}
\end{equation}
with $\tilde{\vartheta}=\partial^{\alpha}\tilde{v}_{\alpha}$ being the (Newtonian) peculiar volume scalar and $\delta=\delta\rho/\rho$ representing the familiar density contrast. Given the profound difference between the above and its relativistic counterpart (compare to Eq.~(\ref{laux1}) in \S~\ref{ssCRM}), one should expect a considerable difference in the conclusions of the two treatments as well. Indeed, combining (\ref{appA1}), (\ref{appA2}) and (\ref{appA4}) leads to~\cite{TKA}
\begin{equation}
\tilde{q}= q+ {2\over3}\,{\tilde{\vartheta}\over H}+ {1\over2}\,\delta\simeq q\,,  \label{appA5}
\end{equation}
since $\tilde{\vartheta}/H,\,\delta\ll1$ throughout the linear regime. Therefore, within the limits of Newtonian gravity, the deceleration parameters measured in the Hubble and in the tilted frames coincide for all practical purposes.

The different result of the Newtonian analysis stems from the fact that, in contrast to relativity, the bulk-flow flux does not contribute to the gravitational field.  As a result the Newtonian version of the relativistic Eq.~(\ref{laux1}) is Eq.~(\ref{appA4}). Put another way, the Newtonian approach cannot naturally reproduce the key relativistic equation, which for our purposes is expression (\ref{laux1}). Overall, the reason the two theories arrive at so different results and conclusions is the fundamentally different way they treat the gravitational field and the sources that contribute to it (see also discussion in \S~\ref{ssCRM}).

\section{Changing frames}\label{app2}
Physics is independent of the coordinate system and only the observers' interpretation of the results may differ. So far, we have taken the viewpoint of the tilted observer, living inside a typical galaxy and moving with peculiar velocity $\tilde{v}_a$ relative to our (reference) Hubble frame. In what follows, we will change our perspective and adopt that of an observer following the reference $u_a$-frame, which has peculiar velocity $v_a$ with respect to the tilted frame. Then, the associated (linearised) Lorentz boost reads
\begin{equation}
u_a= \tilde{u}_a+ v_a\,,  \label{appB1}
\end{equation}
with $\tilde{u}_av^a=0$, $v^2=v_av^a\ll1$ and $v_a=-\tilde{v}_a$. This immediately implies that
\begin{equation}
\tilde{\Theta}= \Theta- \vartheta\,,  \label{appB2}
\end{equation}
where $\vartheta={\rm D}^av_a=-\tilde{\rm D}^a\tilde{v}_a= -\tilde{\vartheta}$ to first approximation.\footnote{The linear relation $\vartheta=-\tilde{\vartheta}$ implies that local expansion for the tilted observers appears as (local) contraction to their idealised counterparts and vice versa.} Recall that overdots indicate covariant time-derivatives in the $u_a$-frame. We will also assume that the energy flux and the 4-acceleration vanish in the tilted frame and thus set $\tilde{q}_a=0= \tilde{A}_a$.\footnote{Both the $u_a$ and the $\tilde{u}_a$ frames are allowed to have shear and vorticity (at the linear level), since none of these kinematic quantities is involved in the calculations.} Then, according to (\ref{lrs1}), we have
\begin{equation}
q_a= -\rho v_a \hspace{10mm} {\rm and} \hspace{10mm} A_a= \dot{v}_a+ Hv_a\,,  \label{appB3}
\end{equation}
at the linear level~\cite{M}.

The deceleration parameters in the two frames are still defined by Eqs.~(\ref{qs}a) and (\ref{qs}b), which combined lead to the linear expression
\begin{equation}
\tilde{q}= q- {\dot{\vartheta}\over3\dot{H}}\left(1+{1\over2}\,\Omega\right)= q+ {\dot{\vartheta}\over3H^2}\,,  \label{appB4}
\end{equation}
since $\vartheta/H\ll1$ throughout the linear regime. Note that the above is equivalent to Eq.~(\ref{tq2}) in \S~\ref{ssTDPs}, once the linear relations $\vartheta=-\tilde{\vartheta}$ and $\dot{\vartheta}=-\tilde{\vartheta}^{\prime}$ are accounted for.

Employing the commutation law between temporal and spatial covariant derivatives (e.g.~see~\cite{TCM,EMM}), we have
\begin{equation}
\dot{\vartheta}= -H\vartheta+ {\rm D}^a\dot{v}_a\,,  \label{appB5}
\end{equation}
to first approximation. In addition, linearising Eq.~(2.3.1) of~\cite{TCM}, or Eq.~(10.101) of~\cite{EMM}, around the reference $u_a$-frame (while keeping in mind that $q_a=-\rho v_a$ there) gives
\begin{equation}
\dot{\Delta}_a= -\mathcal{Z}_a- 3aH\dot{v}_a- 3aH^2v_a+ a{\rm D}_a\vartheta\,.  \label{aaB6}
\end{equation}
Taking the spatial divergence of the above and solving for ${\rm D}^a\dot{v}_a$, we obtain
\begin{equation}
{\rm D}^a\dot{v}_a= -H\vartheta+ {1\over3H}\,{\rm D}^2\vartheta- {1\over3a^2H}\left(\dot{\Delta}+\mathcal{Z}\right)\,,  \label{aaB7}
\end{equation}
which substituted into the right-hand side of (\ref{appB5}) leads to
\begin{equation}
\dot{\vartheta}= -2H\vartheta+ {1\over3H}\,{\rm D}^2\vartheta- {1\over3a^2H}\left(\dot{\Delta}+\mathcal{Z}\right)\,.  \label{appB8}
\end{equation}
Harmonically decomposing the latter and substituting the result into the right-hand side of Eq.~(\ref{appB4}), we arrive at
\begin{eqnarray}
\tilde{q}_{(n)}&=& q- {2\over3}\left[1+{1\over6} \left({\lambda_H\over\lambda_n}\right)^2\right] {\vartheta_{(n)}\over H} \nonumber\\ &&-{1\over9}\left({\lambda_H\over\lambda_K}\right)^2 \left({\dot{\Delta}\over H}+{\mathcal{Z}\over H}\right)\,.  \label{appB9}
\end{eqnarray}
Given that $(\lambda_H/\lambda_K)^2=|\Omega-1|$ and since $\Omega\rightarrow1$ according to the observations, the last term on the right-hand side of above is negligible. As a result, we have
\begin{equation}
\tilde{q}_{(n)}= q- {1\over9} \left({\lambda_H\over\lambda_n}\right)^2{\vartheta_{(n)}\over H}\,,  \label{appB10}
\end{equation}
on subhorizon scales (where $\lambda_n\ll\lambda_H$). Finally, recalling that $\vartheta=-\tilde{\vartheta}$, we deduce that Eq.~(\ref{appB10}) is identical to expression (\ref{tq4}) derived in \S~\ref{ssS-DC}.

Therefore, as expected, the physics remains unaffected by the frame choice. The only difference may be in the interpretation the observers may give to their results. As seen by observers following the reference $u_a$-frame, the deceleration parameter ($\tilde{q}$) measured in the tilted frame drops below $q$, or even becomes negative, when $\vartheta>0$. Recall that, as viewed by the tilted observers, this happens when $\tilde{\vartheta}<0$ (see \S~\ref{sPJL} earlier). This difference of perspectives simply reflects the fact that $\vartheta=-\tilde{\vartheta}$. Put another way, a bulk flow that expands locally in one frame appears to contract in the other and vice versa.

\end{document}